\documentclass{ifae}
\usepackage{epsfig}
\usepackage{axodraw}
\newcommand{\bit}{\begin{itemize}}
\newcommand{\eit}{\end{itemize}}

\newcommand{\be}{\begin{equation}}
\newcommand{\ee}{\end{equation}}
\newcommand{\bc}{\begin{center}}
\newcommand{\ec}{\end{center}}
\newcommand{\nn}{\nonumber}
\newcommand{\f}{\frac}
\newcommand{\ba}{\begin{array}}
\newcommand{\ea}{\end{array}}

\def\gappeq{\mathrel{\rlap {\raise.5ex\hbox{$>$}}
{\lower.5ex\hbox{$\sim$}}}}
\def\lappeq{\mathrel{\rlap{\raise.5ex\hbox{$<$}}
{\lower.5ex\hbox{$\sim$}}}}

\newcommand{\mc}{\mathcal}

\def\beq{\begin{equation}}
\def\eeq{\end{equation}}
\def\bea{\begin{eqnarray}}
\def\eea{\end{eqnarray}}

\newcommand{\cp}{\tilde{\chi}^+_1}
\newcommand{\cm}{\tilde{\chi}^-_1}
\newcommand{\ccp}{\tilde{\chi}^+_2}
\newcommand{\ccm}{\tilde{\chi}^-_2}
\newcommand{\mcu}{m_{\tilde{\chi}^+_1}}
\newcommand{\mcd}{m_{\tilde{\chi}^+_2}}
\newcommand{\eepm}{e^+e^-}
\newcommand{\hcc}{e^+e^-\to h\; \tilde{\chi}^+_1\tilde{\chi}^-_1}
\newcommand{\hccc}{ h\; \tilde{\chi}^+_1\tilde{\chi}^-_1}
\newcommand{\sqs}{\sqrt{s}}
\newcommand{\tb}{\tan{\beta}}
\newcommand{\msne}{m_{\tilde{\nu_e}}}
\newcommand{\sne}{\tilde{\nu_e}}
\newcommand{\maa}{m_{A^0}}
\newcommand{\mh}{m_h}
\newcommand{\dsdh}{E_h\frac{d\sigma}{d^3\mathbf{h}}}

\begin{document}

%--- Header (title, authors, abstract)
\begin{header}
  % The title of your talk
  \title{ Higgs-boson coupling to charginos in the MSSM at linear colliders.}

  % List of authors with affiliations
  \begin{Authlist}
    G.~Ferrera\IIref{uni}{infn}\Acknow{a}, B.~Mele\IIref{infn}{uni}\Acknow{b}

  \Affiliation{uni}{Universit\`a di Roma ``La Sapienza'', Rome, Italy}
  \Affiliation{infn}{INFN, Sezione di Roma, Rome, Italy} 
%  \Affiliation{cern}{CERN, Geneva, Switzerland}
  \Acknowfoot{a}{Talk presented by G.Ferrera, e-mail: giancarlo.ferrera@roma1.infn.it}
  \Acknowfoot{b}{e-mail: barbara.mele@roma1.infn.it}
  \end{Authlist}

  % Collaboration (if applicable)
%  \collaboration {On Behalf of the XYZ Experiment}
  
  % Abstract
  \begin{abstract}
We discuss the associated production of a light Higgs boson ($h$) and a light chargino ($\cp$) pair in
the process 
$\hcc$ in the Minimal Supersymmetric Standard Model (MSSM)
at linear colliders (LC) with $\sqs=500~GeV$.
This process gives direct informations about Higgs-boson coupling to 
light charginos which cannot be analyzed in decay processes
due to phase-space restriction.
We compute  total cross sections in the regions of the
MSSM parameter space where the process $\hcc$ cannot proceed
via  on-shell production and subsequent decay of either heavier charginos
or pseudoscalar Higgs bosons $A$. Cross sections up to a few
$0.1 fb$ are allowed, making this process potentially detectable at 
high-luminosity LC.
We also compute analytically  the final $h$ momentum distributions
in the limit of heavy 
electron-sneutrino masses, $\msne \gg M_W$.
%(or, equivalently to our aims, of dominant higgsino component 
%in the light charginos),
 \end{abstract} 
  
\end{header}

% Beging of the text (no page break)
\section{Introduction}
In this talk we present the main features of the work discussed more 
exhaustively in~\cite{Ferrera:2004hh}.
We know that linear colliders with $\sqs=500~GeV$ would be a very powerful 
{\it precision instrument}
for Higgs-boson physics and physics beyond the standard model (SM) 
that could show up at the LHC. In fact, if  supersymmetry (SUSY) 
exists with partners
of known particles  with masses not too far from present experimental
limits, it will be necessary to study the details of new physics in order to 
understand which SUSY scenario is effectively realized.
Next-generation linear colliders such as TESLA 
and NLC/JLC~\cite{Accomando:1997wt}
would be able to measure (sometimes with excellent precision) 
a number of crucial
parameters (such as masses, couplings and mixing angles),
and eventually test the fine structure of a particular SUSY model.

We know that the Higgs couplings to particle are strictly related to the 
mass of particles. In the case of light particle the Higgs couplings
are suppressed 
(as for the light fermions couplings to the Higgs bosons, 
where $g_{hf\bar f}\sim  m_f/v$) and the coupling can generally be 
determined through the corresponding 
Higgs decay branching ratio measurement.

On the other hand, the Higgs couplings to vector bosons
are unsuppressed and the analysis of the main Higgs-boson production 
cross section, that occurs through the couplings to vector bosons,
can provide a good determination of such couplings. 

There are a number of Higgs-boson couplings 
to quite heavy particles, other than gauge bosons,
that can not be investigated through Higgs-boson decay channels
due to phase-space restrictions.
In this case, 
the associated production of a Higgs boson and a pair of
the heavy particles, when allowed by phase space, can provide
an alternative to measure  the
corresponding coupling, even if 
some reduction  in the rate due to the 
possible phase-space saturation is expected.

For instance, the SM Higgs-boson {\it unsuppressed} coupling to the top quark,
 $ m_t/v$,
can be determined at linear colliders with $\sqs\sim 1$TeV through
the production rates for the Higgs radiated off a top-quark pair
in the channel $e^+e^-\to h \; t \bar t$
\cite{Gaemers:1978jr}.

Our purpose is to use the latter strategy  
in the context of the  MSSM, that introduces  an entire spectrum of 
relatively heavy partners, in many cases  
coupled to Higgs bosons via an unsuppressed coupling constant.
A typical example is that of the light Higgs-boson coupling to the
light top squark $h \;\tilde t_1 \tilde t_1$, that can be naturally large.
The continuum production $e^+e^-\to h \;\tilde t_1 \tilde t_1$ has been
studied in \cite{Belanger:1998rq} as a means of determining this coupling
(the corresponding channel at hadron colliders has been investigated
also in \cite{Djouadi:1997xx}). The Higgs coupling to the $\tau$ slepton
has been considered in~\cite{Datta:2001sh}.

Here, we discuss the possibility to measure
the light Higgs coupling to light charginos $\hccc$ through  the 
Higgs-boson production in association with a chargino pair
at linear colliders through the process 
\beq
\hcc \; .
\label{duno}
\eeq
We will not include in our study the case where the 
considered process
proceeds through the on-shell production of 
either the heavier chargino $\tilde{\chi}^-_2$ or the
pseudoscalar Higgs boson $A$ with a subsequent decay
$\tilde{\chi}^-_2 \to h \cm$ and $A\to\cp\cm$, respectively.
We will also assume a large value for the electron sneutrino mass
(i.e., $\msne>1$ TeV). This suppresses the Feynman diagrams with a
sneutrino exchange, involving predominantly the gaugino
components of the charginos.
Note  that while
 {\it heavy} Higgs bosons couplings to SUSY partners can be mostly explored
 via  Higgs decay rates. 
the {\it light} Higgs-boson coupling to light charginos 
cannot be investigated through  
Higgs decay channels due to  phase-space restrictions.
Indeed, in the MSSM $m_h$ is expected to be lighter that about 
130 GeV \cite{Brignole:2001jy},
and the present experimental limit on the chargino mass
 $m_{\cp}>103.5$ GeV 
 (or even the milder one $m_{\cp}>92.4$ GeV, in case of almost degenerate
 chargino and lightest neutralino)
 \cite{LEPSUSYWG} excludes the decay $h\to\cp\cm$.

Note that the SM process $\eepm \to H W^+W^-$  
(that can be connected by a SuSy
transformation to $\hcc$)  has a total 
cross section of about 5.6 fb  for $m_H\simeq 120$ GeV,
at $\sqs\simeq 500$ GeV \cite{Baillargeon:1993iw}.

%%%%%%%%%%%%%%%%%%%%%%%%%%%%%%
%        DIAGRAMMI
%%%%%%%%%%%%%%%%%%%%%%%%%%%%%%%
\unitlength=1.0 pt
\SetScale{1.0}
\SetWidth{0.7}      % line    size control
%\tiny    %  letter  size control
%{} \qquad\allowbreak
%\hspace{1cm}
%
%  diagram # 1
\begin{figure}[ht!]
\begin{center}
\begin{picture}(160,130)(0,0)
\Text(13.0,90.0)[r]{$e^+$}
\ArrowLine(55,60)(15,90) 
\Text(40.0,82.0)[]{$p_1$}
\Text(13.0,30.0)[r]{$e^-$}
\ArrowLine(15,30)(55,60) 
\Text(40.0,38.0)[]{$p_2$}
\Text(72.0,55.0)[t]{$\gamma$}
\Photon(55.0,60.0)(100.0,60.0){3}{4} 
\Text(72.0,69.0)[]{$k$}
\Text(142.0,30.0)[l]{$\tilde\chi^-_1$}
\ArrowLine(140.0,30.0)(100.0,60.0) 
\Text(120.0,38.0)[]{$q_2$}
\Text(98.0,75.0)[r]{$\tilde\chi^+_1$}
\ArrowLine(100.0,60.0)(100.0,90.0) 
\Text(108.0,75.0)[]{$q_3$}
\Text(142.0,110)[l]{$\tilde\chi^+_1$}
\ArrowLine(100.0,90.0)(140.0,110)
\Text(120.0,107.0)[]{$q_1$}

\Text(142.0,70.0)[l]{$h^0$}
\DashLine(100.0,90.0)(140.0,70.0){2.5} 
\Text(120.0,87.0)[]{$h$}
\Text(77.5,0)[b] {$A_1$}
\end{picture} \ 
%
%{} \qquad\allowbreak
\hspace{2cm}
%
%  diagram # 2
\begin{picture}(160,130)(0,0)
\Text(13.0,90.0)[r]{$e^+$}
\ArrowLine(55,60)(15,90) 
\Text(40.0,82.0)[]{$p_1$}
\Text(13.0,30.0)[r]{$e^-$}
\ArrowLine(15,30)(55,60) 
\Text(40.0,38.0)[]{$p_2$}
\Text(72.0,55.0)[t]{$\gamma$}
\Photon(55.0,60.0)(100.0,60.0){3}{4} 
\Text(72.0,69.0)[]{$k$}
\Text(141.0,90.0)[l]{$\tilde\chi^+_1$}
\ArrowLine(100,60)(140,90) 
\Text(120.0,83.0)[]{$q_2$}
\Text(98.0,45.0)[r]{$\tilde\chi^-_1$}
\ArrowLine(100.0,30.0)(100.0,60.0) 
\Text(108.0,45.0)[]{$q_4$}
\Text(142.0,50.0)[l]{$h^0$}
\DashLine(100.0,30.0)(140.0,50.0){2.5} 
\Text(120.0,34.0)[]{$h$}
\Text(142.0,10)[l]{$\tilde\chi^-_1$}
\ArrowLine(140,10)(100,30)
\Text(120.0,13.0)[]{$q_1$}
\Text(77.5,0)[b] {$A_2$}
\end{picture} \ 
%{} \qquad\allowbreak
\newline
% diagram # 3
%\hspace{1cm}
\begin{picture}(160,130)(0,0)
\Text(13.0,90.0)[r]{$e^+$}
\ArrowLine(55,60)(15,90) 
\Text(40.0,82.0)[]{$p_1$}
\Text(13.0,30.0)[r]{$e^-$}
\ArrowLine(15,30)(55,60) 
\Text(40.0,38.0)[]{$p_2$}
\Text(72.0,55.0)[t]{$Z^0$}
\Photon(55.0,60.0)(100.0,60.0){3}{4} 
\Text(72.0,69.0)[]{$k$}
\Text(142.0,30.0)[l]{$\tilde\chi^-_1$}
\ArrowLine(140.0,30.0)(100.0,60.0) 
\Text(120.0,38.0)[]{$q_2$}
\Text(98.0,75.0)[r]{$\tilde\chi^+_1$}
\ArrowLine(100.0,60.0)(100.0,90.0) 
\Text(108.0,75.0)[]{$q_3$}
\Text(142.0,110)[l]{$\tilde\chi^+_1$}
\ArrowLine(100.0,90.0)(140.0,110)
\Text(120.0,107.0)[]{$q_1$}
\Text(142.0,70.0)[l]{$h^0$}
\DashLine(100.0,90.0)(140.0,70.0){2.5} 
\Text(120.0,87.0)[]{$h$}
\Text(77.5,0)[b] {$A_3$}
\end{picture} \ 
%
%{} \qquad\allowbreak
\hspace{2cm}
%  diagram # 4
\begin{picture}(160,130)(0,0)
\Text(13.0,90.0)[r]{$e^+$}
\ArrowLine(55,60)(15,90) 
\Text(40.0,82.0)[]{$p_1$}
\Text(13.0,30.0)[r]{$e^-$}
\ArrowLine(15,30)(55,60) 
\Text(40.0,38.0)[]{$p_2$}
\Text(72.0,55.0)[t]{$Z^0$}
\Photon(55.0,60.0)(100.0,60.0){3}{4} 
\Text(72.0,69.0)[]{$k$}
\Text(141.0,90.0)[l]{$\tilde\chi^+_1$}
\ArrowLine(100,60)(140,90) 
\Text(120.0,83.0)[]{$q_2$}
\Text(98.0,45.0)[r]{$\tilde\chi^-_1$}
\ArrowLine(100.0,30.0)(100.0,60.0) 
\Text(108.0,45.0)[]{$q_4$}
\Text(142.0,50.0)[l]{$h^0$}
\DashLine(100.0,30.0)(140.0,50.0){2.5} 
\Text(120.0,34.0)[]{$h$}
\Text(142.0,10)[l]{$\tilde\chi^-_1$}
\ArrowLine(140,10)(100,30)
\Text(120.0,13.0)[]{$q_1$}
\Text(77.5,0)[b] {$A_4$}
\end{picture} \ 
%{} \qquad\allowbreak
\newline
%\hspace{1cm}
% diagram # 5
\begin{picture}(160,130)(0,0)
\Text(13.0,90.0)[r]{$e^+$}
\ArrowLine(55,60)(15,90) 
\Text(40.0,82.0)[]{$p_1$}
\Text(13.0,30.0)[r]{$e^-$}
\ArrowLine(15,30)(55,60) 
\Text(40.0,38.0)[]{$p_2$}
\Text(72.0,55.0)[t]{$Z^0$}
\Photon(55.0,60.0)(100.0,60.0){3}{4} 
\Text(72.0,69.0)[]{$k$}
\Text(142.0,30.0)[l]{$\tilde\chi^-_1$}
\ArrowLine(140.0,30.0)(100.0,60.0) 
\Text(120.0,38.0)[]{$q_2$}
\Text(98.0,75.0)[r]{$\tilde\chi^+_2$}
\ArrowLine(100.0,60.0)(100.0,90.0) 
\Text(108.0,75.0)[]{$q_3$}
\Text(142.0,110)[l]{$\tilde\chi^+_1$}
\ArrowLine(100.0,90.0)(140.0,110)
\Text(120.0,107.0)[]{$q_1$}
\Text(142.0,70.0)[l]{$h^0$}
\DashLine(100.0,90.0)(140.0,70.0){2.5} 
\Text(120.0,87.0)[]{$h$}
\Text(77.5,0)[b] {$A_5$}
\end{picture} \ 
%
%{} \qquad\allowbreak
\hspace{2cm}
%  diagram # 6
\begin{picture}(160,130)(0,0)
\Text(13.0,90.0)[r]{$e^+$}
\ArrowLine(55,60)(15,90) 
\Text(40.0,82.0)[]{$p_1$}
\Text(13.0,30.0)[r]{$e^-$}
\ArrowLine(15,30)(55,60) 
\Text(40.0,38.0)[]{$p_2$}
\Text(72.0,55.0)[t]{$Z^0$}
\Photon(55.0,60.0)(100.0,60.0){3}{4} 
\Text(72.0,69.0)[]{$k$}
\Text(141.0,90.0)[l]{$\tilde\chi^+_1$}
\ArrowLine(100,60)(140,90) 
\Text(120.0,83.0)[]{$q_2$}
\Text(98.0,45.0)[r]{$\tilde\chi^-_2$}
\ArrowLine(100.0,30.0)(100.0,60.0) 
\Text(108.0,45.0)[]{$q_4$}
\Text(142.0,50.0)[l]{$h^0$}
\DashLine(100.0,30.0)(140.0,50.0){2.5} 
\Text(120.0,34.0)[]{$h$}
\Text(142.0,10)[l]{$\tilde\chi^-_1$}
\ArrowLine(140,10)(100,30)
\Text(120.0,13.0)[]{$q_1$}
\Text(77.5,0)[b] {$A_6$}
\end{picture} \ 
%
%{} \qquad\allowbreak
\newline
%\hspace{1cm}
%  diagram # 7
\hspace{2cm}
\begin{picture}(160,130)(0,0)
\Text(13.0,90.0)[r]{$e^+$}
\ArrowLine(55,60)(15,90) 
\Text(40.0,82.0)[]{$p_1$}
\Text(13.0,30.0)[r]{$e^-$}
\ArrowLine(15,30)(55,60) 
\Text(40.0,38.0)[]{$p_2$}
\Text(72.0,55.0)[t]{$Z^0$}
\Photon(55.0,60.0)(100.0,60.0){3}{4} 
\Text(72.0,69.0)[]{$k$}
\Text(141.0,90.0)[l]{$h^0$}
\DashLine(100,60)(140,90){2.5} 
\Text(120.0,83.0)[]{$h$}
\Text(98.0,45.0)[r]{$Z^0$}
\Photon(100.0,30.0)(100.0,60.0){3}{3} 
\Text(107.0,45.0)[]{$q$}
\Text(142.0,50.0)[l]{$\tilde\chi+_1$}
\ArrowLine(100.0,30.0)(140.0,50.0) 
\Text(120.0,33.0)[]{$q_2$}
\Text(142.0,10)[l]{$\tilde\chi^-_1$}
\ArrowLine(140,10)(100,30)
\Text(120.0,13.0)[]{$q_1$}
\Text(77.5,0)[b] {$A_7$}
\end{picture} \ 
%{} \qquad\allowbreak
\caption{{\small
Set of Feynman diagrams included in the analytic Higgs-boson distribution.
 }}
\label{uno}
\end{center}
\end{figure}
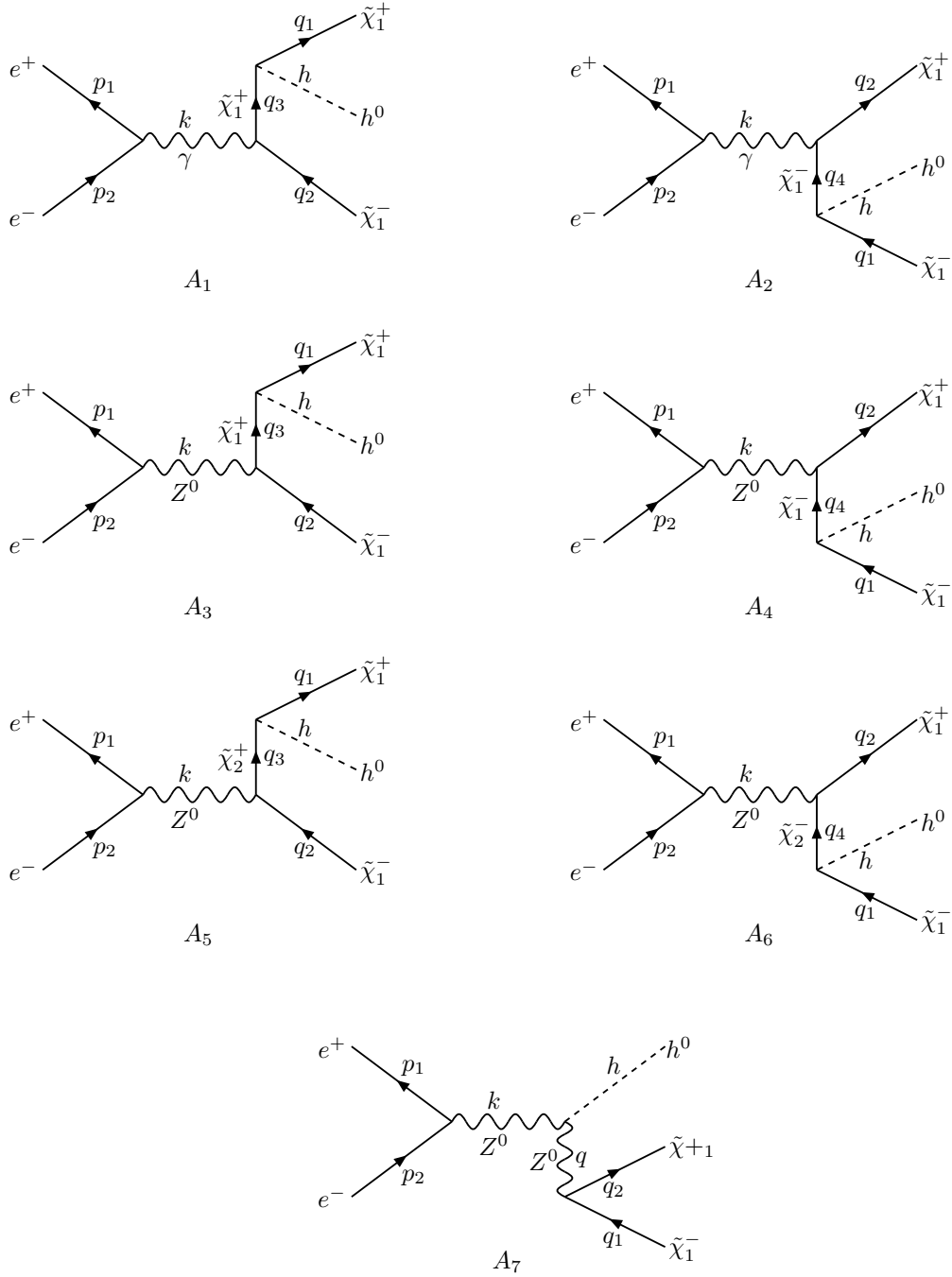
\begin{figure}[ht!]
%\vspace{2cm}{\normalsize I diagrammi di Feynmann trascurati sono:}
%\hspace{1cm}
%\titlepage
\begin{center}
%  diagram # 8
\begin{picture}(160,130)(0,0)
\Text(13.0,90.0)[r]{$e^+$}
\ArrowLine(15.0,90.0)(55.0,60.0) 
\Text(13.0,30.0)[r]{$e^-$}
\ArrowLine(55.0,60.0)(15.0,30.0) 
\Text(77,64.0)[b]{$Z^0$}
\Photon(55.0,60.0)(100.0,60.0){3}{4} 
\Text(141.0,90.0)[l]{$h^0$}
\DashLine(100,60)(140,90){2.5} 
\Text(98.0,45.0)[r]{$A^0$}
\DashLine(100.0,30.0)(100.0,60.0){2.5} 
\Text(142.0,50.0)[l]{$\tilde\chi^+_1$}
\ArrowLine(100.0,30.0)(140.0,50.0) 
\Text(142.0,10)[l]{$\tilde\chi^-_1$}
\ArrowLine(140,10)(100,30)
\Text(77.5,0)[b] {$A_8$}
\end{picture}
{} \qquad\allowbreak
\hspace{2cm}
%  diagram # 9
%\begin{figure}[h]
%\begin{center}
\begin{picture}(160,130)(0,0)
\Text(13.0,90.0)[r]{$e^+$}
\ArrowLine(55,90)(15,90) 
\Text(13.0,30.0)[r]{$e^-$}
\ArrowLine(15,30)(55,30) 
\Text(51,60)[r]{$\tilde\nu$}
\DashArrowLine(55,30)(55,90){2.5}
\Text(142,90)[l]{$\tilde\chi^+_1$}
\ArrowLine(55,90)(140,90)
\Text(77.5,33)[b]{$\tilde\chi^+_1$}
\ArrowLine(100,30)(55,30)
\Text(142.0,50.0)[l]{$h^0$}
\DashLine(100.0,30.0)(140.0,50.0){2.5} 
\Text(142.0,10)[l]{$\tilde\chi^-_1$}
\ArrowLine(140,10)(100,30)
\Text(77.5,0)[b] {$A_9$}
\end{picture} \ 
\newline
%  diagram # 10
\begin{picture}(160,130)(0,0)
\Text(13.0,90.0)[r]{$e^+$}
\ArrowLine(55,90)(15,90) 
\Text(13.0,30.0)[r]{$e^-$}
\ArrowLine(15,30)(55,30) 
\Text(51,60)[r]{$\tilde\nu$}
\DashArrowLine(55,30)(55,90){2.5}
\Text(142,90)[l]{$\tilde\chi^+_1$}
\ArrowLine(55,90)(140,90)
\Text(77.5,33)[b]{$\tilde\chi^+_2$}
\ArrowLine(100,30)(55,30)
\Text(142.0,50.0)[l]{$h^0$}
\DashLine(100.0,30.0)(140.0,50.0){2.5} 
\Text(142.0,10)[l]{$\tilde\chi^-_1$}
\ArrowLine(140,10)(100,30)
\Text(77.5,0)[b] {$A_{10}$}
\end{picture} \
%\end{center}
%\end{figure}
%{} \qquad\allowbreak
%\hspace{1cm}
{} \qquad\allowbreak
\hspace{2cm}
%  diagram # 11
%\begin{figure}[h]
%\begin{center}
\begin{picture}(160,130)(0,0)
\Text(13.0,90.0)[r]{$e^+$}
\ArrowLine(55,90)(15,90) 
\Text(13.0,30.0)[r]{$e^-$}
\ArrowLine(15,30)(55,30) 
\Text(51,60)[r]{$\tilde\nu$}
\DashArrowLine(55,30)(55,90){2.5}
\Text(142,30)[l]{$\tilde\chi^-_1$}
\ArrowLine(140,30)(55,30)
\Text(77.5,93)[b]{$\tilde\chi^+_1$}
\ArrowLine(55,90)(100,90)
\Text(142.0,110.0)[l]{$h^0$}
\DashLine(100.0,90.0)(140.0,110.0){2.5} 
\Text(142.0,70)[l]{$\tilde\chi^+_1$}
\ArrowLine(100,90)(140,70)
\Text(77.5,0)[b] {$A_{11}$}
\end{picture} \ 
\newline 
%  diagram # 12
\begin{picture}(160,130)(0,0)
\Text(13.0,90.0)[r]{$e^+$}
\ArrowLine(55,90)(15,90) 
\Text(13.0,30.0)[r]{$e^-$}
\ArrowLine(15,30)(55,30) 
\Text(51,60)[r]{$\tilde\nu$}
\DashArrowLine(55,30)(55,90){2.5}
\Text(142,30)[l]{$\tilde\chi^-_1$}
\ArrowLine(140,30)(55,30)
\Text(77.5,93)[b]{$\tilde\chi^+_2$}
\ArrowLine(55,90)(100,90)
\Text(142.0,110.0)[l]{$h^0$}
\DashLine(100.0,90.0)(140.0,110.0){2.5} 
\Text(142.0,70)[l]{$\tilde\chi^+_1$}
\ArrowLine(100,90)(140,70)
\Text(77.5,0)[b] {$A_{12}$}
\end{picture} \ 
%\end{center}
%\end{figure}
%{} \qquad\allowbreak
%\hspace{1cm}
{} \qquad\allowbreak
\hspace{2cm}
% diagram # 13
%\begin{figure}[h]
%\begin{center}
\begin{picture}(160,130)(0,0)
\Text(13.0,90.0)[r]{$e^+$}
\ArrowLine(77.5,90)(15,90) 
\Text(13.0,30.0)[r]{$e^-$}
\ArrowLine(15,30)(77.5,30) 
\Text(74,75)[r]{$\tilde\nu$}
\DashArrowLine(77.5,60)(77.5,90){2.5}
\Text(74,45)[r]{$\tilde\nu$}
\DashArrowLine(77.5,30)(77.5,60){2.5}
\Text(142,90)[l]{$\tilde\chi^+_1$}
\ArrowLine(77.5,90)(140,90)
\Text(142,30)[l]{$\tilde\chi^+_1$}
\ArrowLine(140,30)(77.5,30)
\Text(142.0,60)[l]{$h^0$}
\DashLine(77.5,60)(140.0,60){2.5} 
\Text(77.5,0)[b] {$A_{13}$}
\end{picture} \ 
{} \qquad\allowbreak
\caption{{\small 
Set of Feynman diagrams not included in the analytic Higgs-boson distribution.
}}
\label{due}
\end{center}
\end{figure}
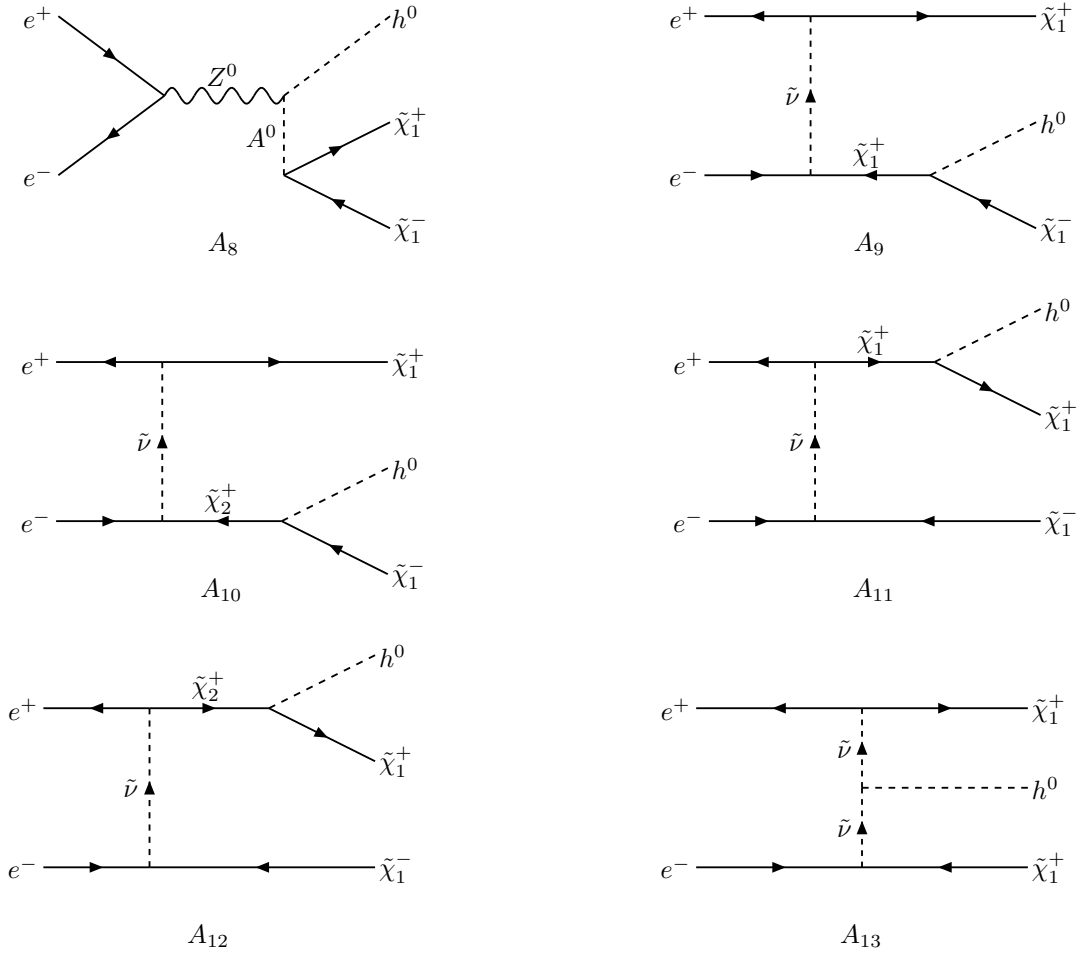

%%%%%%%%%%%%%%%%%%%%%%%%%%%%%%%%%%%%%%%%%%%%%%%
\section{Relevant MSSM Scenarios}
%%%%%%%%%%%%%%%%%%%%%%%%%%%%%%%%%%%%%%%%%%%%%%%%
In the MSSM, charginos are the mass eigenstates of the mass matrix 
that mixes charged gauginos and 
higgsinos (see \cite{haber}, \cite{Ferrera:2004hh}). 
At tree level, the mass eigenvalues $\mcu$ and $\mcd$ and the mixing 
angles can be 
analytically written in terms of the parameters $M_2$, $\mu$ and $\tb$.
The presence of a Higgs boson in the process $\hcc$ requires
a further
parameter, that can be the pseudoscalar mass $\maa$. 
On the other hand, the inclusion of the main radiative corrections to 
the Higgs-boson mass
%%%%%%%%%%%%%%%%%%%%%%%%%%%%%%%%%%%
%%%%%%%%%%%%%%%%%%%%%%%%%%%%%%%%%%%%%
and couplings
involves all the basic parameters needed for 
setting the complete
mass spectrum  of the SuSy partners in the MSSM.
%In our study of $\hcc$ at $\sqs=$500 GeV, 
We set $\maa=500$ GeV, this
pushes the pseudoscalar field $A^0$ beyond the threshold for direct production,
thus preventing resonant $A^0\to \cp\cm$ contribution to the $\hccc$ final
state.
At the same time, this choice for $\maa$ 
sets a {\it decoupling-limit} scenario ($\maa\gg M_Z$). 

Present experimental lower limits
on $\mh$ \cite{lephiggsMSSM} in the {\it decoupling-limit} MSSM  
are close to the ones derived from  the SM Higgs boson direct search
(i.e., $m_H>114.4$ GeV at 95\% C.L. \cite{lephiggsSM}).

The corrections to the light Higgs mass and  coupling parameter 
$\alpha$
have been computed according to the code FeynHiggsFast \cite{FeynHiggsFast}, 
with the following  input
parameters :  
$M_{\tilde{t}_{L,R}}=M_{\tilde{b}_{L,R}}=M_{\tilde{g}}=1$ TeV , 
$X_t\;(\equiv A_t-\mu \cot{\beta})=$ either 0 or 2 TeV,
$A_b=A_t$, $m_t=175$ GeV, $m_b=4.5$ GeV, $\mu=200$ GeV, $M_2=400$ GeV, 
and renormalization scale at $m_t$, in the most complete version 
of the code
(varying the $\mu$ and $M_2$
parameters 
would affect the Higgs spectrum and couplings negligibly).

We assumed  three different $\tb$ scenarios, 
and corresponding $\mh$ values for $\maa=500$ GeV:

\vspace{0.2cm}
\noindent
{\bf a)} $\tb=3$, with {\it maximal} stop mixing 
(i.e., $X_t=2$ TeV), 
and $\mh=120.8$ GeV;

\noindent
{\bf b)} $\tb=15$, with {\it no} stop mixing (i.e., $X_t=0$), 
and $\mh=114.3$ GeV;

\noindent
{\bf c)} $\tb=30$, with {\it maximal} stop mixing (i.e., $X_t=2$ TeV), 
and $\mh=132.0$ GeV; 

\vspace{0.2cm}
\noindent
that are allowed by present experimental limits \cite{lephiggsMSSM}.
In this talk we focalize on the scenario {\bf a)}.

\begin{figure}[th]
\vspace{0cm}
\centerline{\epsfxsize=8.0truein \epsfbox{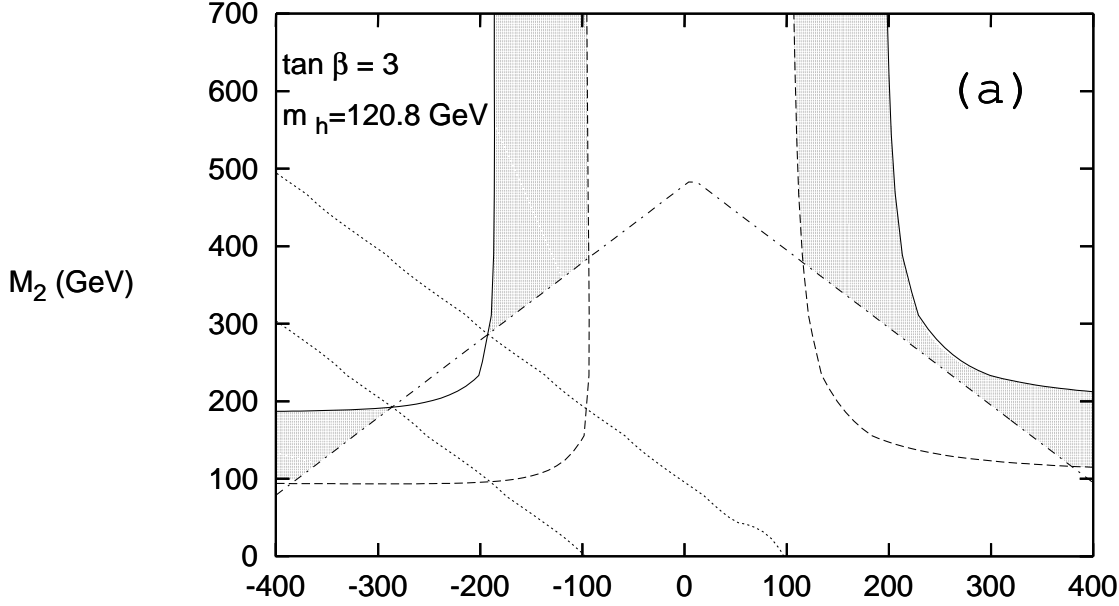}}
\vspace{-2cm}
\caption{
\small 
Parameter regions allowed for the continuum production
$\hcc$ at $\sqrt s=500~GeV$, for $\tan\beta=3$ and $M_{A^0}=500~GeV$
(in grey). 
}
\label{tre}
\vspace{0.5cm}
\end{figure}

There are 13 Feynman diagrams involved in the process $\hcc$,
7 with  
the $s$-channel $Z^0$/$\gamma$ exchange and 6 with the $t$-channel 
electron-sneutrino $\sne$ exchange. 
In our cross-section evaluation, we include only  the $s$-channel 
diagrams reported in 
Fig.~\ref{uno}, and disregard  the 6 diagrams in Fig.~\ref{due}. In fact 
the latter are
expected to contribute moderately to the cross section in the case
$\msne> 1$ TeV, $\maa=500$ GeV. We discuss the accuracy of
this assumption in~\cite{Ferrera:2004hh}.
In Fig.~\ref{tre}, we show (in grey), 
the area in the $(\mu,M_2)$ plane that is of relevance for the 
{\it non resonant}
$\hcc$ process.
The solid lines correspond to the threshold energy contour level :
\beq
\sqs=2\; \mcu +\mh ,
\label{sogliauno}
\eeq 
while the dashed lines  refer to the experimental limit on the 
light chargino mass ($\mcu\simeq 100$ GeV).

The straight dot-dashed lines limit from above the region that allows
the associated production of a light chargino $\cp$ and
a resonant heavier chargino $\ccm$
 (that we are not interested in), and correspond to :
\beq
\sqs=\mcu +m_{\ccm} .
\label{sogliadue}
\eeq  
A further region of interest (beyond the grey one) could be the one
where, although $\sqs>\mcu +m_{\ccm}$, the heavier chargino is {\it below}
the threshold for a direct decay $\ccp \to \cp h$.
Then, again a resonant $\ccp$  would not  be allowed.
The  area where $\mcd<\mcu+\mh$ is the one inside the oblique
stripes in Fig.~\ref{tre}.
The intersection of these stripes with the area between the
solid and dashed curves is a further (although quite restricted) region
relevant to the non resonant $\hcc$ process. 
In our analysis, 
we did not include this parameter region, 
since this would have required 
some further
dedicated elaboration of the analytic form for the 
$\hcc$ distributions. 

%%%%%%%%%%%%%%%%%%%%%%%%%%%%%%%%%%%%%
%%%%%%%%%%%%%%%%%%%%%%%%%%%%%%%%%%%%%%%%%%%%%%%%%%%%%%%%%%%%%%%%%%%%%%
%\clearpage
%%%%%%%%%%%%%%%%%%%%%%%%%%%%%%%%%%%%
\section{Cross Sections and Distributions}
As anticipated, our analysis  concentrates  
on the set of 7 Feynman diagrams presented in Fig.~\ref{uno}.
Our evaluation will then be particularly suitable in case of heavy 
electron sneutrinos. %We will discuss quantitatively this point 
%in Section 4.

The matrix elements corresponding to the amplitudes $A_1,\dots,A_7$ 
in Fig.~\ref{uno} are :
\bea
%M1
{\mc M_1}&=& \f{ige^2}{k^2+i\epsilon}\bar u^\chi_{s_1}(q_1)
\left(C^L_{11}P_L+C^R_{11}P_R\right)
\f{(\not{\!q_3}+M_1)}{q_3^2-M_1^2+i\epsilon}\gamma^\mu v^\chi_{s_2}(q_2)
\bar v^e_{r_1}(p_1)
\gamma_\mu u^e_{r_2}(p_2)\nn \\
%M2
{\mc M_2}&=& \f{ige^2}{k^2+i\epsilon}\bar u^\chi_{s_1}(q_1) \gamma^\mu
\f{(-\!\!\not{\!q_4}+M_1)}{q_4^2-M_1^2+i\epsilon}
\left(C^L_{11}P_L+C^R_{11}P_R\right) v^\chi_{s_2}(q_2)
\bar v^e_{r_1}(p_1)
\gamma_\mu u^e_{r_2}(p_2)\nn \\
%M3
{\mc M_3}&=& \f{-ig^3}{4\cos^2\theta_w (k^2-M_Z^2+i\epsilon)}
\bar u^\chi_{s_1}(q_1)\left(C^L_{11}P_L+C^R_{11}P_R\right)
\f{(\not{\!q_3}+M_1)}{q_3^2-M_1^2+i\epsilon}\gamma^\mu \nn \\
&& \times\left( O^L_{11}P_L+O^R_{11} P_R \right) v^\chi_{s_2}(q_2)
\left( g_{\mu\nu}-\f{k_\mu k_\nu}{M_Z^2}\right) \bar v^e_{r_1}(p_1)
\gamma^\nu(g_V-\gamma_5) u^e_{r_2}(p_2)\nn \\
%M4
{\mc M_4}&=&\f{-ig^3}{4\cos^2\theta_w (k^2-M_Z^2+i\epsilon)}
\bar u^\chi_{s_1}(q_1)\gamma^\mu
\left( O^L_{11}P_L+O^R_{11} P_R \right) \f{(-\!\!\not{\!q_4}+M_1)}
{q_4^2-M_1^2+i\epsilon}\nn \\
&&\times\left(C^L_{11}P_L+C^R_{11}P_R\right) v^\chi_{s_2}(q_2)
\left( g_{\mu\nu}-\f{k_\mu k_\nu}{M_Z^2}\right) \bar v^e_{r_1}(p_1)
\gamma^\nu (g_V-\gamma_5)u^e_{r_2}(p_2)\nn \\
%M5
{\mc M_5}&=& \f{-ig^3}{4\cos^2\theta_w (k^2-M_Z^2+i\epsilon)}
\bar u^\chi_{s_1}(q_1)\left(C^L_{12}P_L+C^R_{12}P_R\right)
\f{(\not{\!q_3}+M_2)}{q_3^2-M_2^2+i\epsilon}\gamma^\mu \nn \\
&& \times\left( O^L_{21}P_L+O^R_{21} P_R \right) v^\chi_{s_2}(q_2)
\left( g_{\mu\nu}-\f{k_\mu k_\nu}{M_Z^2}\right) \bar v^e_{r_1}(p_1)
\gamma^\nu(g_V-\gamma_5) u^e_{r_2}(p_2)\nn \\
%M6
{\mc M_6}&=&\f{-ig^3}{4\cos^2\theta_w (k^2-M_Z^2+i\epsilon)}
\bar u^\chi_{s_1}(q_1)\gamma^\mu
\left( O^L_{12}P_L+O^R_{12} P_R \right) \f{(-\!\!\not{\!q_4}+M_2)}
{q_4^2-M_2^2+i\epsilon}\nn \\
&&\times\left(C^L_{21}P_L+C^R_{21}P_R\right) v^\chi_{s_2}(q_2)
\left( g_{\mu\nu}-\f{k_\mu k_\nu}{M_Z^2}\right) \bar v^e_{r_1}(p_1)
\gamma^\nu(g_V-\gamma_5) u^e_{r_2}(p_2)\nn \\
%M7
{\mc M_7}&=&\f{ig^3M_Z\sin{(\beta-\alpha)}}{4\cos^3\theta_w}
\bar u^\chi_{s_1}(q_1)\gamma^\mu
\left( O^L_{11}P_L+O^R_{11} P_R \right) v^\chi_{s_2}(q_2)\nn \\
&&\times \f{(g_{\mu\nu}-q_\mu q_\nu /M_Z^2)}
{(q^2-M_Z^2+i\epsilon)} \f{(g^{\nu\sigma}-k^\nu k^\sigma /M_Z^2)}
{(k^2-M_Z^2+i\epsilon)}\bar v^e_{r_1}(p_1)
\gamma_\sigma (g_V-\gamma_5) u^e_{r_2}(p_2) ,
\label{matrix}
\vspace{0.2cm}
\eea
where  
$$
\;\;\;k=p_1+p_2=q_1+q_2+h, \;\;\;\;\;q_3=q_1+h, \;\;\;\;\;q_4=q_2+h, 
\;\;\;\;\;q=p_1+p_2-h, 
$$
and
$\;\;M_{1,2}=m_{\tilde\chi_{1,2}^\pm}$.

All external momenta are defined in Fig.~\ref{uno}, 
as flowing from the left to the right, 
and different couplings in Eq.~(\ref{matrix}) are defined 
in~\cite{Ferrera:2004hh}. The lower indices of the spinors $u,v$ refer to the 
particle spin.

We squared, averaged over the initial spin, and summed over the final spin
the sum of the matrix elements in Eq.~(\ref{matrix}) with the help of FORM
\cite{form}.
Then, we performed a double analytic integration over 
the phase-space variables.
This allowed us to obtain an exact {\it analytic} expression for the Higgs-boson 
momentum distribution
\beq
  E_h\frac{d\sigma}{d^3\mathbf{h}}=
  \f{\beta}{s(4\pi)^5} \int_{-1}^1 d\cos\vartheta \; \int_0^{2\pi}d\varphi\;
  |\overline{\mathcal{M}}|^2\;= f(p_1,p_2,h)\; ,
\label{dh}
\eeq
 
where $\mathcal{M}
=\sum^7_{i=1}\mathcal{M}_i\;$.
The complete code, including the analytic result for $\dsdh$ 
(that is a quite lengthy expression),
and the numerical integration routine that allows a fast
evaluation of the total cross section, is available from the authors' e-mail
addresses.

%%%%%%%%%%%%%%%%%%%%%%%%%%%%%%%%%%%%%%%%%%%%%%%%%%%%%%%%%%%%%%%%%%%%%%
\begin{figure}[th]
\vspace{0cm}
\centerline{\epsfxsize=8.0truein \epsfbox{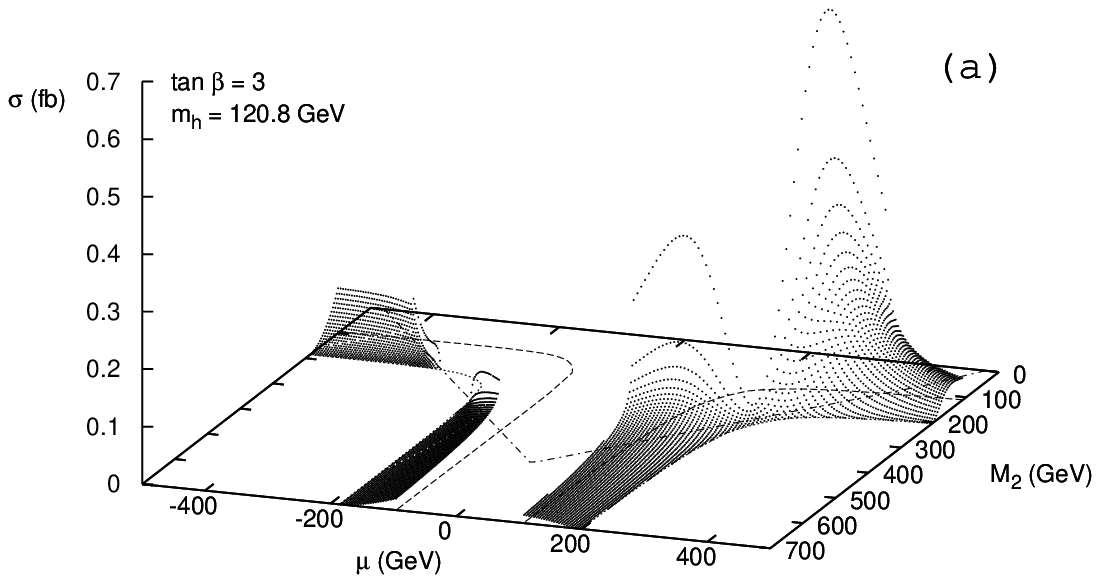}}
\vspace{-0.5cm}
\caption{
\small
Total cross section for
$\hcc$ at $\sqrt s=500~GeV$, for $\tan\beta=3$ and $M_{A^0}=500~GeV$. 
}
\label{tot}
\vspace{0.5cm}
\end{figure}
%\clearpage
In Fig.~\ref{tot}, we show the total cross section for
$\hcc$ at $\sqrt s=500~GeV$. We obtained it  by numerically integrating
over the Higgs-boson energy and angle
the analytic distribution in Eq.~(\ref{dh}).
In Fig.~\ref{tot}, we scan the relevant $(\mu, M_2)$ parameter space
for $|\mu|<500$ GeV, and $M_2<700$ GeV. Cross sections of a few $0.1 fb^{-1}$
are reached in a good portion of the allowed regions especially 
for positive $\mu$.
We checked that our results completely agree with the cross section 
evaluated by CompHEP
\cite{comphep} on the basis of the same
set of Feynman diagrams of Fig.~\ref{uno}.

We studied quantitatively the consequence
 of disregarding the 6 diagrams in Fig.~\ref{due},
involving either pseudoscalar or sneutrino  exchange, by comparing
our results with the cross section corresponding to  the complete set of
 13 diagrams, computed by CompHEP. 
While the pseudoscalar-exchange diagram never
contributes sizable for $\maa=500$ GeV, the influence of 
the 5 sneutrino-exchange
diagrams depends critically on $\msne$ and also
on the relative importance of
the gaugino-higgsino
components in the chargino~\cite{Ferrera:2004hh}.

%%%%%%%%%%%%%%%%%%%%%%%%%%%%%%%%%%%%%%%%%%%%%%%%%%%%%%%%%%%%
%%%%%%%%%%%%%%%%%%%%%%%%%%%%%%%%%%%%%%%%%%%%%%%%%%%%%%%%%%%%%%%%%%%%%%
\begin{figure}[th]
\vspace{-0cm}
\centerline{\epsfxsize=4.truein \epsfbox{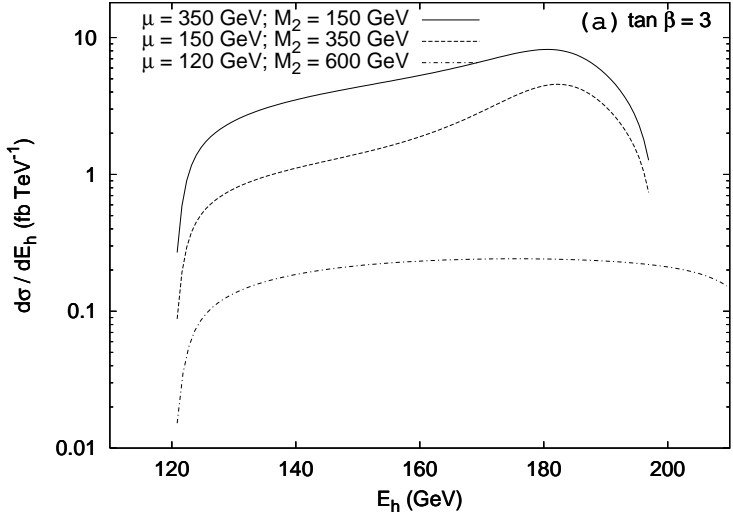}}
\caption{
\small
Higgs-boson energy distribution in the $\eepm$ c.m. frame in
$\hcc$ at $\sqrt s=500~GeV$, for $\tan\beta=3$, $M_{A^0}=500~GeV$.
}
\label{dsde}
\vspace{0.5cm}
\end{figure}

%%%%%%%%%%%%%%%%%%%%%%%%%%%%%%%%%%%%
%%%%%%%%%%%%%%%%%%%%%%%%%%%%%%%%%%%%
%%%%%%%%%%%%%%%%%%%%%%%%%%%%%%%%%%%%%%%%%%%%%%%%%%%%%%%%%%%%%%%%%%%%%%
\begin{figure}[th]
\vspace{-0cm}
\centerline{\epsfxsize=4.truein \epsfbox{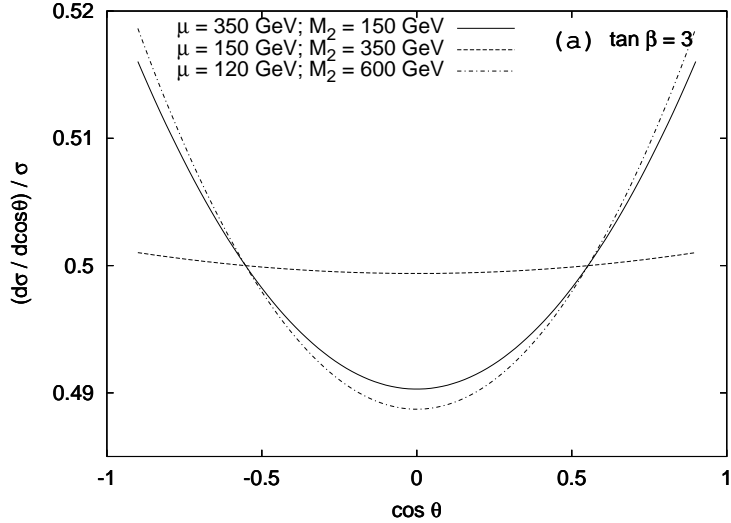}}
\caption{
\small
Higgs-boson angular distribution in the $\eepm$ c.m. frame in 
$\hcc$ at $\sqrt s=500~GeV$, for $\tan\beta=3$, $M_{A^0}=500~GeV$.
}
\label{dsdc}
\vspace{0.5cm}
\end{figure}
%\clearpage
We also considered the $\hcc$ cross sections at the higher-energy
extension expected for a linear-collider project \cite{Accomando:1997wt}.
Going at $\sqs\simeq 1$ TeV, our treatment of the $\hcc$ cross section
becomes less accurate.
As far as production rates are concerned,
for chargino and Higgs-boson masses not much
heavier than present experimental limits, and heavy sneutrinos,
the $\sqrt s=500~GeV$ phase
of the linear collider could be the best option
to study the process $\hcc$.
%%%%%%%%%%%%%%%%%%%%%%%%%%%%%%%%%%%%

%%%%%%%%%%%%%%%%%%%%%%%%%%%%%%%%%%%%
%\section{Higgs-boson Momentum Distributions}

We finally studied the behavior of Higgs-boson energy and angular distributions
versus the MSSM parameters.
In Figs.~\ref{dsde} and \ref{dsdc}, we plot energy and angular distributions in 
the $\eepm$ c.m. frame, as obtained by numerically integrating over one variable
Eq.~(\ref{dh}).
Both the
energy- and angular-distribution shapes are 
considerably influences by the
gaugino/higgsino composition of the light chargino, and 
by  a possible saturation of
the available phase-space~\cite{Ferrera:2004hh}.

\section{Conclusions}
We analyzed the associated
production of a light Higgs 
boson  and a light-chargino pair  
in the MSSM  at $\sqs=500~GeV$ $e^+e^+$ linear colliders. 
The process was discussed in the region of the MSSM parameter space where
there is no resonant production of either 
the heavier chargino or the pseudoscalar 
Higgs boson $A$ and in the limit of heavy sneutrino masses
($\msne > 1$ TeV). 
We computed analytically distributions 
in the Higgs-boson momentum. 
We obtained cross sections up to a few $0.1~fb$ (for $\mu>0$)
for masses not much heavier than
present experimental limits.
These rates make this process potentially detectable at 
$\sqrt s=500~GeV$ $\eepm$ LC
with an integrated luminosity of the order of
$1000 fb^{-1}$. 
This could allow a first determination of the 
$\hccc$ production rate with a statistical
error of the order of $10\%$,
that, in absence of further systematics, could be extrapolated
to a determination of the $\hccc$ coupling with  comparable accuracy.
Future studies in this direction are requires for a more solid assessment
of the potential of this process.
%\vspace{2cm}
\newpage
%%%%%%%%%%%%%%%%%%%%%%%%%%%%%%%%%%%%


\begin{thebibliography}{20}

%\cite{Ferrera:2004hh}
\bibitem{Ferrera:2004hh}
G.~Ferrera and B.~Mele,
%``Associated production of a light Higgs boson and a chargino pair in the MSSM
%at linear colliders,''
arXiv:hep-ph/0406256.
%%CITATION = HEP-PH 0406256;%%


%\cite{Accomando:1997wt}
\bibitem{Accomando:1997wt}
E.~Accomando {\it et al.}  [ECFA/DESY LC Physics Working Group
                  Collaboration],
%``Physics with e+ e- linear colliders'',
Phys.\ Rept.\  {\bf 299} (1998) 1.
[arXiv:hep-ph/9705442]; \\
%%CITATION = HEP-PH 9705442;%%
%\cite{Aguilar-Saavedra:2001rg}
%\bibitem{Aguilar-Saavedra:2001rg}
J.~A.~Aguilar-Saavedra {\it et al.}  [ECFA/DESY LC Physics Working Group
                  Collaboration],
%``TESLA Technical Design Report Part III: Physics at an e+e- Linear 
%Collider'',
arXiv:hep-ph/0106315; \\
%%CITATION = HEP-PH 0106315;%%
%\cite{Abe:2001gc}
%\bibitem{Abe:2001gc}
K.~Abe {\it et al.}  [ACFA Linear Collider Working Group Collaboration],
%``Particle physics experiments at JLC'',
arXiv:hep-ph/0109166; \\
%%CITATION = HEP-PH 0109166;%%
%\cite{Abe:2001nn}
%\bibitem{Abe:2001nn}
T.~Abe {\it et al.}  [American Linear Collider Working Group Collaboration],
%``Linear collider physics resource book for Snowmass 2001. 1:  Introduction'',
in {\it Proc. of the APS/DPF/DPB Summer Study on the Future of Particle 
Physics (Snowmass 2001) } ed. N.~Graf,
arXiv:hep-ex/0106055 ;
arXiv:hep-ex/0106056 ;
arXiv:hep-ex/0106057 ; 
arXiv:hep-ex/0106058.
%%CITATION = HEP-EX 0106055;%%



%\cite{Gaemers:1978jr}
\bibitem{Gaemers:1978jr}
K.~J.~Gaemers and G.~J.~Gounaris,
%``Bremsstrahlung Production Of Higgs Bosons In E+ E- Collisions'',
Phys.\ Lett.\ B {\bf 77} (1978) 379; \\
%%CITATION = PHLTA,B77,379;%%
%\cite{Djouadi:gp}
%\bibitem{Djouadi:gp}
A.~Djouadi, J.~Kalinowski and P.~M.~Zerwas,
%``Measuring The H T Anti-T Coupling In E+ E- Collisions'',
Mod.\ Phys.\ Lett.\ A {\bf 7} (1992) 1765 and
Z.\ Phys.\ C {\bf 54} (1992) 255; \\
%%CITATION = MPLAE,A7,1765;%%
%\cite{Gunion:1996vv}
%\bibitem{Gunion:1996vv}
J.~F.~Gunion, B.~Grzadkowski and X.~G.~He,
%``Determining the top anti-top and Z Z couplings of a neutral Higgs 
% boson of arbitrary CP nature at the NLC'',
Phys.\ Rev.\ Lett.\  {\bf 77} (1996) 5172,
[arXiv:hep-ph/9605326] ; \\
%%CITATION = HEP-PH 9605326;%%
%\cite{Baer:1999ge}
%\bibitem{Baer:1999ge}
H.~Baer, S.~Dawson and L.~Reina,
%``Measuring the top quark Yukawa coupling at a linear e+ e- collider'',
Phys.\ Rev.\ D {\bf 61} (2000) 013002,
[arXiv:hep-ph/9906419] ; \\
%%CITATION = HEP-PH 9906419;%%
%\cite{Juste:1999af}
%\bibitem{Juste:1999af}
A.~Juste and G.~Merino,
%``Top-Higgs Yukawa coupling measurement at a linear e+ e- collider'',
arXiv:hep-ph/9910301 ; \\
%%CITATION = HEP-PH 9910301;%%
%\cite{Moretti:1999kx}
%\bibitem{Moretti:1999kx}
S.~Moretti,
%``The process e+ e- $\to$ H t anti-t and its backgrounds at future  
%electron positron colliders'',
Phys.\ Lett.\ B {\bf 452} (1999) 338,
[arXiv:hep-ph/9902214] ; \\
%%CITATION = HEP-PH 9902214;%%
%\cite{Denner:2003zp}
%\bibitem{Denner:2003zp}
A.~Denner, S.~Dittmaier, M.~Roth and M.~M.~Weber,
%``Radiative corrections to Higgs-boson production in association with 
%top-quark pairs at e+ e- colliders'',
arXiv:hep-ph/0309274 and references therein.
%%CITATION = HEP-PH 0309274;%%


%\cite{Belanger:1998rq}
\bibitem{Belanger:1998rq}
G.~Belanger, F.~Boudjema, T.~Kon and V.~Lafage,
%``Associated stop Higgs production at the linear collider and 
%extraction  of the stop parameters'',
Eur.\ Phys.\ J.\ C {\bf 9} (1999) 511,
[arXiv:hep-ph/9811334] ; \\
%%CITATION = HEP-PH 9811334;%%
%\cite{Djouadi:1999dg}
%\bibitem{Djouadi:1999dg}
A.~Djouadi, J.~L.~Kneur and G.~Moultaka,
%``Associated production of Higgs bosons with scalar quarks at future  
%hadron and e+ e- colliders'',
Nucl.\ Phys.\ B {\bf 569} (2000) 53,
[arXiv:hep-ph/9903218].
%%CITATION = HEP-PH 9903218;%%


%\cite{Djouadi:1997xx}
\bibitem{Djouadi:1997xx}
A.~Djouadi, J.~L.~Kneur and G.~Moultaka,
%``Higgs boson production in association with scalar top quarks at 
%proton  colliders'',
Phys.\ Rev.\ Lett.\  {\bf 80} (1998) 1830,
[arXiv:hep-ph/9711244].
%%CITATION = HEP-PH 9711244;%%
%\cite{Dedes:1998yt}
%\bibitem{Dedes:1998yt}
A.~Dedes and S.~Moretti,
%``Pseudoscalar Higgs production in association with stop and sbottom  
% pairs at the LHC in the MSSM'',
Phys.\ Rev.\ D {\bf 60} (1999) 015007 
[arXiv:hep-ph/9812328] and Eur.\ Phys.\ J.\ C {\bf 10} (1999) 515.
[arXiv:hep-ph/9904491] ; \\
%%CITATION = HEP-PH 9812328;%%
%``Higgs production in association with squark pairs in the minimal  
%supersymmetric standard model at future hadron
%colliders'',
%\cite{Belanger:1999pv}
%\bibitem{Belanger:1999pv}
G.~Belanger, F.~Boudjema and K.~Sridhar,
%``SUSY Higgs at the LHC: Large stop mixing effects and associated  
%production'',
Nucl.\ Phys.\ B {\bf 568} (2000) 3,
[arXiv:hep-ph/9904348].
%%CITATION = HEP-PH 9904348;%%

%\cite{Datta:2001sh}
\bibitem{Datta:2001sh}
A.~Datta, A.~Djouadi and J.~L.~Kneur,
%``Probing the SUSY Higgs boson couplings to scalar leptons at high-energy  e+
%e- colliders,''
Phys.\ Lett.\ B {\bf 509} (2001) 299
[arXiv:hep-ph/0101353].
%%CITATION = HEP-PH 0101353;%%

%\cite{Brignole:2001jy}
\bibitem{Brignole:2001jy}
A.~Brignole, G.~Degrassi, P.~Slavich and F.~Zwirner,
%``On the O(alpha(t)**2) two-loop corrections to the neutral Higgs 
%boson  masses in the MSSM'',
Nucl.\ Phys.\ B {\bf 631} (2002) 195
[arXiv:hep-ph/0112177] and
%%CITATION = HEP-PH 0112177;%%
%``On the two-loop sbottom corrections to the neutral Higgs boson 
% masses  in the MSSM'',
Nucl.\ Phys.\ B {\bf 643} (2002) 79
[arXiv:hep-ph/0206101].
%%CITATION = HEP-PH 0206101;%%


%\cite{LEPSUSYWG}
\bibitem{LEPSUSYWG}
LEPSUSYWG, ALEPH, DELPHI, L3 and OPAL experiments, note LEPSUSYWG/01-03.1
and note LEPSUSYWG/02-04.1
(http://lepsusy.web.cern.ch/lepsusy/Welcome.html). 


\bibitem{Baillargeon:1993iw}
M.~Baillargeon, F.~Boudjema, F.~Cuypers, E.~Gabrielli and B.~Mele,
 %``Higgs production in association with a vector boson pair at future e+ e-
%colliders'',
Nucl.\ Phys.\ B {\bf 424} (1994) 343
[arXiv:hep-ph/9307225].

\bibitem{haber}
H.~E.~Haber and G.~L.~Kane,
%``The Search For Supersymmetry: Probing Physics Beyond The Standard Model'',
Phys.\ Rept.\  {\bf 117} (1985) 75 .
%%CITATION = PRPLC,117,75;%%
%%%CITATION = HEP-PH 0402295;%%
\bibitem{lephiggsMSSM} 
ALEPH, DELPHI, L3, OPAL Collaborations and the LEP Higgs Working Group,
LHWG Note 2001-4 {\it [ALEPH 2001-057, DELPHI 2001-114, L3 Note 2007, OPAL 
Technical Note TN699]}, CERN preprint 2001; \\
OPAL Collaboration, OPAL PN524, CERN preprint 2003.

\bibitem{lephiggsSM}
%\bibitem{Barate:2003sz}
R.~Barate {\it et al.} 
[ALEPH, DELPHI, L3, OPAL
Collaborations and LEP Working Group for Higgs boson searches],
%``Search for the standard model Higgs boson at LEP'',
Phys.\ Lett.\ B {\bf 565} (2003) 61
[arXiv:hep-ex/0306033].

\bibitem{FeynHiggsFast}
%\bibitem{Heinemeyer:2000nz}
S.~Heinemeyer, W.~Hollik and G.~Weiglein,
 ``FeynHiggsFast: A program for a fast calculation of masses and 
  mixing  angles in the Higgs sector of the MSSM'',
arXiv:hep-ph/0002213.

\bibitem{form}
J.A.M.~Vermaseren, {\it Symbolic Manipulation with FORM},
published by CAN (Computer Algebra Nederland), Kruislaan 413, 1098
SJ Amsterdam, 1991, ISBN 90-74116-01-9.

\bibitem{comphep}
A.~Pukhov {\it et al.},
``CompHEP: A package for evaluation of Feynman diagrams and integration  over
multi-particle phase space. User's manual for version 33'',
arXiv:hep-ph/9908288; \\
A.~Semenov,
``CompHEP/SUSY package'',
Nucl.\ Instrum.\ Meth.\ A {\bf 502} (2003) 558
[arXiv:hep-ph/0205020].


\end{thebibliography}
\end{document}